\documentclass[preprint,showpacs,preprintnumbers,amsmath,amssymb]{revtex4}

\usepackage{dcolumn}
\usepackage{bm}
\usepackage[pdftex]{graphicx}

\def\etal{{\em et al.}}

\newcommand{\be}{\begin{equation}}
\newcommand{\ee}{\end{equation}}
\newcommand{\ba}{\begin{eqnarray}}
\newcommand{\ea}{\end{eqnarray}}

\def\micron{$\mu$m}
\newcommand{\Nqp}{N_{\rm qp}}

\begin{document}

\preprint{Mazin/aSi}

\title{Thin film dielectric microstrip kinetic inductance detectors}

\author{Benjamin A. Mazin}
\email{bmazin@physics.ucsb.edu}
\homepage{http://www.physics.ucsb.edu/~bmazin/}

\author{Daniel Sank}
\author{Sean McHugh}
\author{Erik A. Lucero}
\author{Andrew Merrill}
\affiliation{Department of Physics, University of California, Santa Barbara, CA 93106-9530}

\author{Jiansong Gao}
\author{David Pappas}
\affiliation{National Institute of Standards and Technology, Boulder, CO 80305-3328}

\author{David Moore}
\author{Jonas Zmuidzinas}
\affiliation{Department of Physics, California Institute of Technology, Pasadena, CA 91125}

\date{\today}


\begin{abstract}
Microwave Kinetic Inductance Detectors, or MKIDs, are a type of low temperature detector that exhibit intrinsic frequency domain multiplexing at microwave frequencies.  We present the first theory and measurements on a MKID based on a microstrip transmission line resonator. A complete characterization of the dielectric loss and noise properties of these resonators is performed, and agrees well with the derived theory.  A competitive noise equivalent power of 5$\times$10$^{-17}$ W Hz$^{-1/2}$ at 1 Hz has been demonstrated.  The resonators exhibit the highest quality factors known in a microstrip resonator with a deposited thin film dielectric.
\end{abstract}



\maketitle

\begin{figure}
\includegraphics[width=0.5\columnwidth]{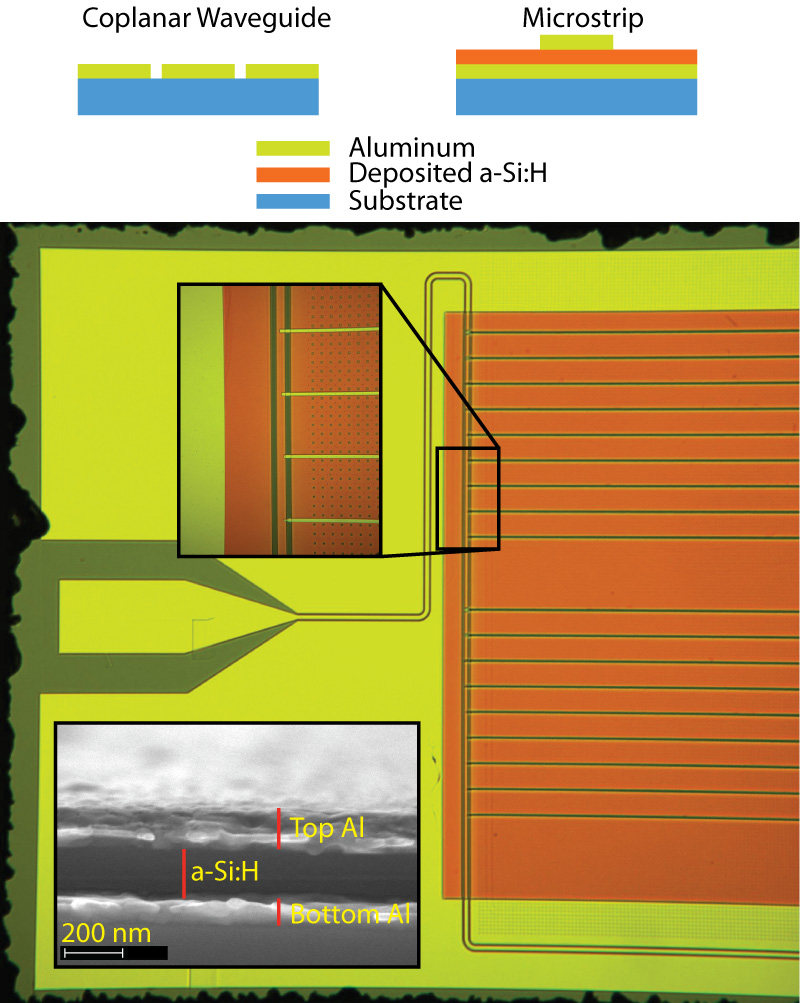}
\caption{(Color online) Top: A cross sectional view of a conventional CPW transmission line and the microstrip transmission lines used in this paper. Bottom: An optical microscope image of the a portion of the tested device.  The inset on the lower left shows a SEM image of the cross section of a device that has been cleaved.  The inset SEM image was used to determine the dielectric thickness $d=200$ nm, the top aluminum wiring layer thickness of $t_1=154$ nm, and the bottom aluminum wiring layer thickness of $t_2=93$ nm.} 
\label{fig:dev}
\end{figure}

Thin film superconducting microwave resonators have been an area of intense research in the past decade for Microwave Kinetic Inductance Detectors (MKIDs)~\cite{day03} for submillimeter\cite{maloney09}, optical/UV~\cite{moore09}, and X-ray~\cite{Mazin:2006p104} astrophysics, components in superconducting qubits~\cite{Wallraff:2004p2846,Martinis:2005p2578}, and fundamental studies in condensed matter physics such as searches for macroscopic quantum states~\cite{Regal:2008p2877} and measurements of the properties of cryogenic liquids~\cite{Grabovskij:2008p2443}.  These resonators have primarily been coplanar waveguide (CPW) transmission line resonators~\cite{Mazin:2004p2}.  A CPW transmission line is a planar structure with a center strip and slots that separate the center strip from ground planes on either side, as shown in the top panel of Figure~\ref{fig:dev}.  They are simple to fabricate out of a single superconducting film on a crystalline dielectric.  However, in many cases the flexibility of a microstrip resonator, which is a stacked structure with a deposited dielectric separating two conductors, would be advantageous.   A microstrip MKID can be deposited on any material, instead of just single crystal silicon or sapphire, and can be made significantly smaller than a CPW MKID.  This allows more flexibility in detector design. The low loss deposited dielectrics required for a sensitive microstrip MKID have many other uses, such as in the lumped element capacitors and wiring crossovers in superconducting qubits~\cite{Martinis:2009p2935} or the microstrip combiner networks of planar antenna arrays~\cite{Day:2006p121}.  This work details the first microstrip MKID using a deposited thin film dielectric.

Microstrip resonators can be made very sensitive by making the dielectric comparable to or thinner than the penetration depth of the superconductor since this causes the device to be dominated by the kinetic inductance of the superconductor, not the magnetic inductance of the transmission line.  This leads to a very sensitive detector since MKIDs respond to changes in the kinetic inductance, such as those caused by broken Cooper Pairs from photon absorption.  It also makes a compact resonator since the phase velocity on the transmission line can be as low as several percent of the speed of light.  Following Swihart~\cite{Swihart:1961p2749} and Pond \etal~\cite{POND:1987p2646}, the phase velocity of a superconducting microstrip whose width $w$ is much greater than the dielectric thickness $d$ and at $T \ll T_c$ can be written using the two fluid approximation as \be
v_p = c \left[ \epsilon_r \left( 1+\frac{\lambda_1}{d}\coth \left(\frac{t_1}{\lambda_1} \right)  + \frac{\lambda_2}{d}\coth \left( \frac{t_2}{\lambda_2} \right) \right) \right]^{-1/2}
\label{eq:vp}
\ee
where $c$ is the speed of light in vacuum, $\epsilon_r$ is the relative dielectric constant of the microstrip dielectric, and $\lambda$ and $t$ are the penetration depth and the thickness of the top (1) and bottom (2) superconductors. The kinetic inductance fraction, $\alpha = \frac{L_k}{L_T}$, is defined as the ratio of the kinetic inductance $L_k$ to the total inductance, $L_T = L_k + L_m$, where $L_m$ is the magnetic inductance of the transmission line.  Using Equation~\ref{eq:vp} and the phase velocity of a normal metal transmission line $v_{pN}=c/\sqrt{\epsilon_r}$ we can calculate $\alpha$ for a uniform distribution of quasiparticles in both the top and bottom microstrip wiring as:
\begin{eqnarray}
\alpha &=& 1 - \left(\frac{v_p}{v_{pN}}\right)^2 \\ 
&=&1-\left( 1+\frac{\lambda_1}{d}\coth \left(\frac{t_1}{\lambda_1} \right)  + \frac{\lambda_2}{d}\coth \left( \frac{t_2}{\lambda_2} \right) \right)^{-1}
\end{eqnarray}

The microstrip MKID is read out by sending a microwave probe signal past the resonator, and a homodyne mixing scheme is used to recover the phase and dissipation changes imprinted onto the carrier by the MKID~\cite{day03}.  Using Equation 3 and Equation~\ref{eq:resp}~\cite{Gao:2008p421}, we can express the expected responsivity, normalized so the microwave transmission past the resonator $S_{21}$ far off resonance is unity, of a microstrip MKID in both dissipation and phase as:

\be
\frac{\partial S_{21}}{\partial N_{qp}} = \frac{\alpha |\gamma| \kappa Q_m^2}{V Q_c} 
\label{eq:resp}
\ee
with
\begin{eqnarray*}
\kappa \approx \frac{1}{\pi N_0} \sqrt{\frac{2}{\pi k T \Delta_0}} \sinh(\xi) K_0(\xi) \\+ j \frac{1}{2 N_0 \Delta_0} 
\left[ 1 + \sqrt{\frac{2 \Delta_0}{\pi k T}}e^{-\xi} I_0(\xi) \right]
\end{eqnarray*}
where $N_{qp}$ is the number of quasiparticles in the resonator, $Q_m$ is the measured quality factor, $Q_c$ is the coupling quality factor, $V$ is twice the volume of the top microstrip wiring layer since this is where the current flows and where quasiparticles effectively contribute to the surface impedance, $N_0$ is the single spin density of states, $\Delta_0$ is the effective gap at $T\approx0$, $\xi=\hbar \omega / 2 k T$, and $\gamma$ is constant that varies from -1/3 in the extreme anamalous limit to -1 in the thin film local limit.  The predicted phase responsivity in radians per quasiparticle, $\partial \theta / \partial \Nqp$, can be found by taking the imaginary part of Equation~\ref{eq:resp}, while the dissipation response $\partial D / \partial \Nqp$ is found by taking the real part.  $Q_m$ is related to $Q_c$ and the quality factor resulting from any source of dissipation in the system, $Q_i$, by the relation $Q_m^{-1} = Q_c^{-1} + Q_i^{-1}$.  We operate the devices in this paper at $T < T_c/8$, so there are essentially no thermal quasiparticles in the devices.  At these temperatures and in a device in which there is no trapped magnetic flux, $Q_i$ should be dominated by losses in the microstrip dielectric. 

This expression for the responsivity of a microstrip MKID can be combined with the known amplifier noise, providing a closed form expression that can predict the sensitivity of the MKID at a given readout power level.

Previous work with lumped element resonators~\cite{OConnell:2008p2581} has shown that hydrogen rich amorphous silicon (a-Si:H) is a promising material, with a low power loss tangent $\tan(\delta)$$\sim$2$\times$10$^{-5}$.  A low loss tangent, and hence a higher internal quality factor since $\tan(\delta)=1/Q_i$, is vital for microstrip MKIDs. A higher $Q_i$ will directly lead to resonators with higher sensitivity when using a readout scheme based on dissipation.  A readout based on dissipation, and not phase, will likely be required since the two-level system (TLS) contribution to the noise equivalent power (NEP) in the phase direction from the deposited dielectric is expected to be much higher than the NEP in the dissipation direction~\cite{Gao:2008p66,Gao:2008p341}.  

In this work microstrip resonators with aluminum wiring and an a-Si:H dielectric have been fabricated. The geometry is illustrated in Figure~\ref{fig:dev}. The resonators were fabricated with optical lithography and dry etching techniques. First, a 93 nm thick aluminum film is deposited to form a CPW feed line and a ground plane with small ground plane holes to suppress effects from stray magnetic fields. This layer is then patterned with a dry etch in an inductively coupled plasma (ICP) etcher.  A 200 nm thick film of a-Si:H is then deposited at a temperature of 100 C to form the insulating dielectric layer.  To make a quarter wave device, a via could now be etched in the a-Si:H using the ICP to form a short to the bottom wiring layer, while a half wave resonator will have an open end without a via.  A 154 nm thick layer of aluminum is then sputtered. This layer is patterned by an ICP to form the top conductor of the microstrip.  This forms microstrip resonators with a width $w$ of 4 \micron~and lengths between 3.9 and 6.0 mm.  The strength of the coupling $(Q_c)$ of the microstrip to the feedline is determined by the amount of microstrip line that covers the CPW feedline. Finally, the a-Si:H insulating layer is patterned with the ICP to allow easy wire bonding, and the silicon wafer is diced into 7.5$\times$2 mm chips. 


The chips are glued into a gold-plated copper sample box with GE varnish and wire bonded to transition boards which convert from the coaxial input lines to a CPW transmission line.  The box is placed inside an adiabatic demagnetization refrigerator (ADR) capable of reaching base temperatures below 100 mK.  A coaxial feedline drives the device through a 30 dB attenuator at 4 Kelvin, and a high electron mobility (HEMT) amplifier with a noise temperature $T_n\approx5.5$~K is used to boost the output signal.  A cryoperm magnetic shield is used to shield the device, and a Helmholtz Coil internal to the magnetic shield is used to apply a magnetic field normal to the surface of the chip. 

\begin{figure}
\includegraphics[width=0.5\columnwidth]{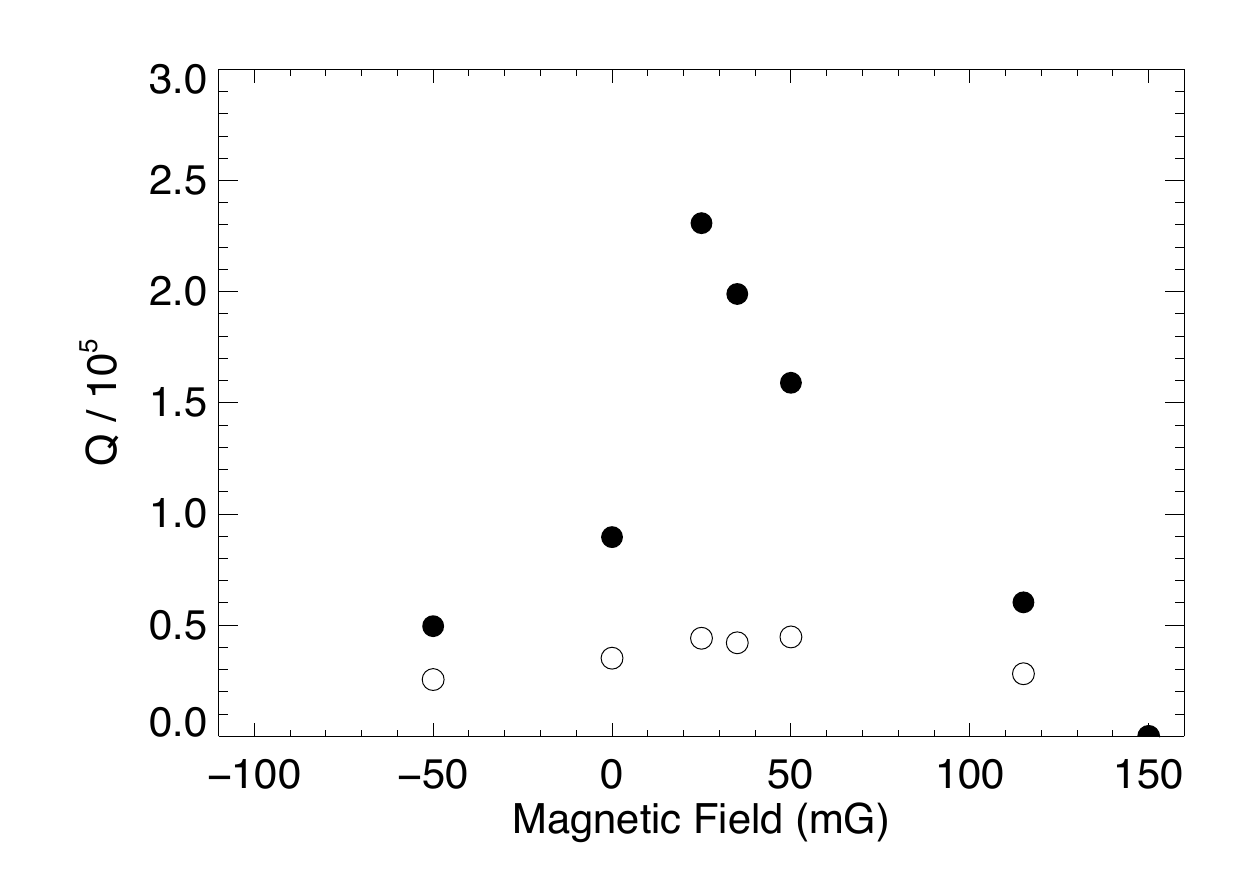}
\caption{The magnetic field dependance of the quality factor of the resonator.  The measured quality factor $Q_m$ is shown as open circles, and the internal quality factor $Q_i$ is shown as filled circles.} 
\label{fig:QvsH}
\end{figure}

Microstrip resonators, like larger CPW resonators\cite{Wang:2009p3037}, appear to be quite sensitive to the magnetic field normal to the metal surface during cooling through the superconducting transition temperature $T_c$.  Figure~\ref{fig:QvsH} shows the applied magnetic field during cooling and the resulting quality factors of the lowest $Q_m$ resonator on the device.  Despite the magnetic shield the best quality factor is achieved with an applied magnetic field of around 30 mG.  This residual field is likely related to the field leaking out of the superconducting magnet in the ADR.  Once cold, the devices showed little response to applied magnetic fields of over several hundred mG, where the resonator $Q_i$ starts degrading.  This degradation remains when the field is ramped down to zero.  When the device was warmed past $T_c$ and recooled the original $Q_i$ was recovered, likely indicating trapped magnetic flux. 

Using the measured thickness of the a-Si:H film of 200 nm from Figure~\ref{fig:dev} and 50 nm as the penetration depth of aluminum, Equation~\ref{eq:vp}, and the known lengths and resonant frequencies of the resonators it is possible to derive the dielectric constant of a-Si:H.  This calculation yields $\epsilon_r = 11.4$.  The error in the determination of $\epsilon_r$ scales with the error in the measurement of the dielectric thickness.  Most of the computed quantities in this paper like NEP vary only weakly with $\epsilon_r$.

\begin{figure}
\includegraphics[width=0.5\columnwidth]{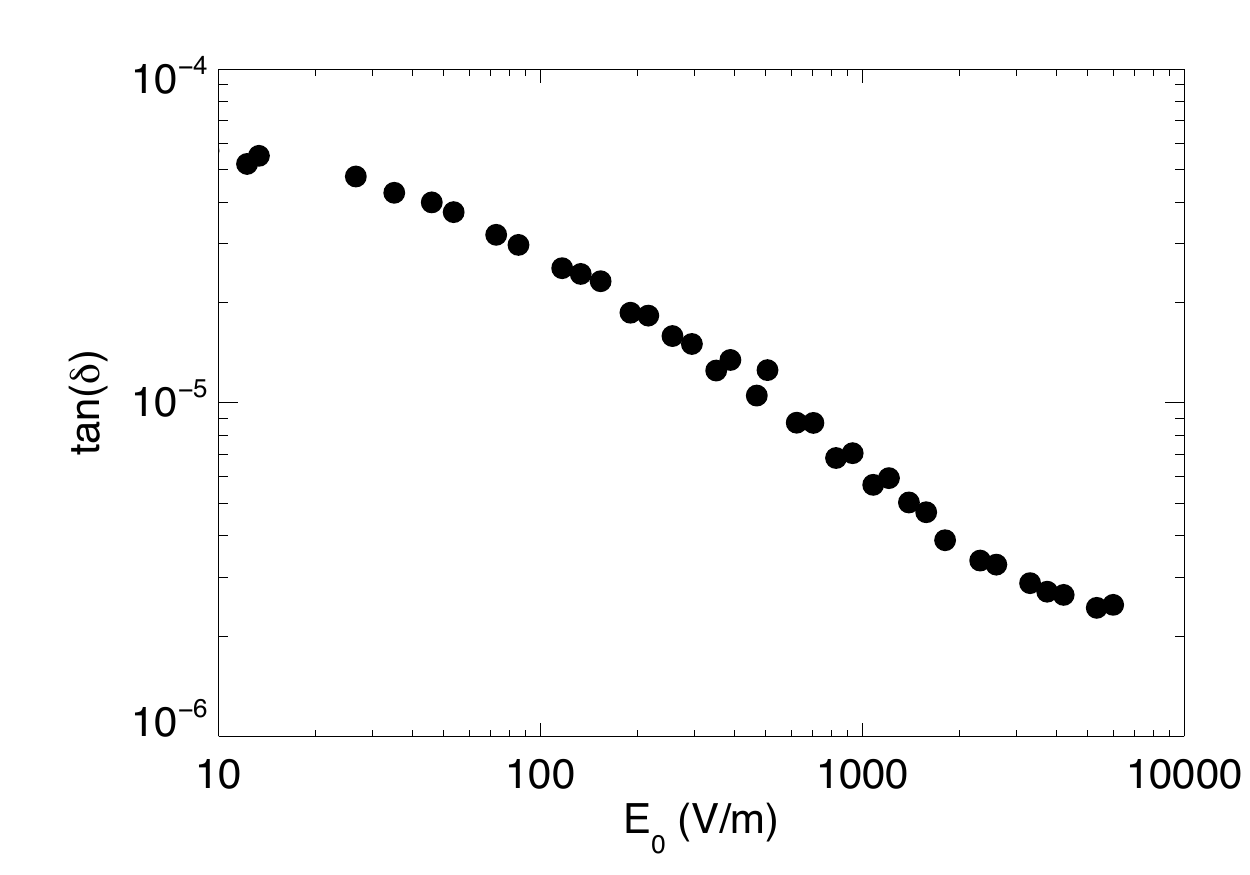}
\caption{The loss tangent of the a-Si:H dielectric, $\tan(\delta)$, dramatically decreases as the electric field in the resonator increases.  The electric field $E_0$ shown is the value at the open ends of the half wave resonator tested.  This resonator is 3985 \micron~ long, with $Q_m=35500$ and a resonant frequency of 9.054 GHz when operated under an optimal magnetic field of 30 mG.  It shows a fractional frequency noise~\cite{Gao:2008p341} of 4.1$\times$10$^{-18}$ Hz$^{-1}$ at a readout power of -91 dBm (equivalent to a current density of 17.5 A/m).} 
\label{fig:QvP}
\end{figure}

Previous work has shown that the dielectric loss tangent is strongly dependent on the electric field in the dielectric~\cite{Martinis:2005p2578,OConnell:2008p2581,Gao:2008p66,Gao:2008p341}.  Figure~\ref{fig:QvP} shows the loss tangent as a function of electric field in one of the a-Si:H microstrip resonators.  The electric field at the open ends of the resonator can be calculated from~\cite{Gao:2008p421}: 
\be
E_0 = \frac{1}{d} \sqrt{\frac{4 Z_0}{\pi m} \frac{Q_m^2}{Q_c} P }
\label{eq:v0}
\ee
where $m=1/2$ for a half wave resonator, $P$ is the microwave power on the feedline, $Q_m$ is the measured quality factor, $Q_c$ is the coupling quality factor, and $Z_0$ is the characteristic impedance of the microstrip transmission line $Z_0 = \sqrt{L/C} = 1/v_p C$.  Since $d \ll w$ we use a parallel plate capacitor approximation for the capacitance per unit length, $C = \epsilon_0 \epsilon_r w /d$.

Since resonators used as detectors will nearly always be operated at the highest readout power possible before non-linear effects set in, the right side of the plot with loss tangents below 2$\times$10$^{-6}$ are the most relevant for MKIDs.  The flattening of the loss tangent towards the right side of the plot is due to the high readout power generating quasiparticles in the resonator.

The sensitivity of the MKID can be calculated by first taking noise spectra on resonance in both the phase $(S_{\theta})$ and dissipation $(S_D)$ direction at a readout power just below the level where the MKID becomes nonlinear.  The dimensionless phase or dissipation shift per quasiparticle referenced to the center of the resonance loop, $\partial \theta/\partial N_{qp}$ and $\partial D/\partial N_{qp}$, can be computed by taking a temperature sweep of the resonance loop and converting the temperature to an effective number of quasiparticles in the resonator center strip, $N_{qp} = 2 N_0 V \sqrt{2 \pi k_B T \Delta_0 } e^{-\Delta_0/k_B T} $.  Performing this operation on the resonator measured in Figure~\ref{fig:QvP} leads to $\partial \theta/\partial N_{qp}=8.2$$\times$$10^{-7}$ and $\partial D/\partial N_{qp}=3.8$$\times$$10^{-7}$ radians per quasiparticle.  Equation~\ref{eq:resp} predicts $\partial \theta/\partial N_{qp}=4.9$$\times$$10^{-7}$ and $\partial D/\partial N_{qp}=1.3$$\times$$10^{-7}$.  The match between the predicted and measured responsivities is reasonably good, showing that the derived responsivity formalism is a reasonably good approximation.

Using the measured noise spectra with this responsivity and a conservative quasiparticle lifetime $\tau_{qp}$ in aluminum of 250~$\mu$s allows us to calculate the NEP$_{\theta}$~\cite{Mazin:2004p2},
\be {\rm
NEP_{\theta}^2}(\omega) = S_\theta(\omega) \left(
                      \frac{\eta \tau_{qp}}{\Delta_0}
                      \frac{\partial \theta}{\partial \Nqp}
                      \right)^{-2} (1 + \omega^2 \tau_{qp}^2)(1 + \omega^2 \tau_{res}^2)
\label{eq:NEP}
\ee
where $\tau_{res} = Q_m/\pi f_0$. This same equation can be used to calculate NEP$_{D}$ by substituting $S_{D}$ for $S_{\theta}$ and $\partial D/\partial N_{qp}$ for $\partial \theta/\partial N_{qp}$.  It can also predict the resonator sensitivity in the dissipation direction by using $\partial D/\partial N_{qp}$ from the real part of Equation~\ref{eq:resp} and the normalized voltage noise of an amplifier with noise temperature $T_n$ referenced to the center of the resonance loop, 
\be
S_D^{pred} = \frac{2 k T_n}{P} \left( \frac{Q_c}{Q_m} \right)^2. 
\label{eq:sd} 
\ee

\begin{figure}
\includegraphics[width=0.5\columnwidth]{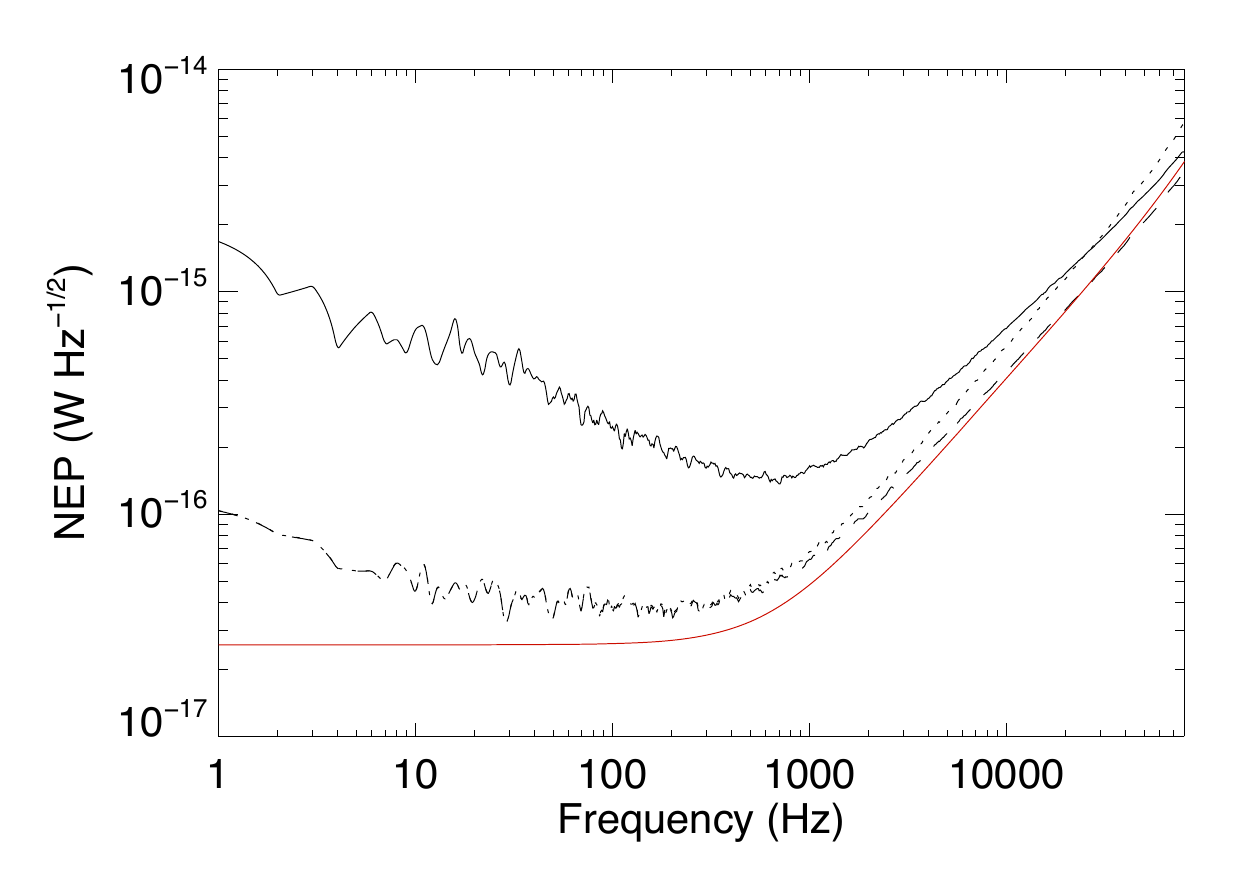}
\caption{(Color online) The noise equivalent power (NEP) of the microstrip resonator.  The solid line shows the NEP derived from phase shifts only, the dotted line is derived from amplitude data, and the dashed line is the optimal NEP using both amplitude and phase data~\cite{Gao:2008p421}.  The red line is the predicted NEP from Equations~\ref{eq:resp},~\ref{eq:NEP},~and~\ref{eq:sd} assuming there is 2 dB of loss between the device and a HEMT amplifier with $T_n=5.5$ K.} 
\label{fig:NEP}
\end{figure}

Combining Equations \ref{eq:resp},~\ref{eq:NEP},~and \ref{eq:sd} yields the predicted NEP based only on the device properties, shown as the red line in Figure~\ref{fig:NEP}.  This is a powerful tool for optimizing microstrip MKIDs for specific detector applications.


Figure~\ref{fig:NEP} shows the calculated NEP for the resonator measured in Figure~\ref{fig:QvP}.  As expected, NEP$_\theta$ is higher than NEP$_D$ due to the contribution of TLSs to the phase noise.  The measured NEP$_D$, with a minimum below 5$\times$10$^{-17}$ W Hz$^{-1/2}$ between 1 Hz and 1 kHz, is a very good NEP competitive with many other low temperature detectors.  There is an small unexplained rise in the NEP$_D$ at very low frequencies, which may be due to incomplete decomposition of the noise into phase and dissipation.  

The ease and flexibility of fabrication combined with the good sensitivity makes microstrip MKIDs an extremely interesting device for future large detector arrays.  Significantly higher performance in much smaller resonators can be achieved by using thinner films and superconductors like titanium that have long penetration depths.

\begin{acknowledgments}
This material is based upon work supported by the National Aeronautics and Space Administration under Grant NNH06ZDA001N-APRA2 issued through the Science Mission Directorate.  The authors would like to thank John Martinis, Sunil Golwala, and Andrew Cleland for useful insights.
\end{acknowledgments}


\end{document}